\documentclass[prl,amssymb,amsmath,showpacs,twoside,twocolumn
]{revtex4}

\usepackage{graphicx}
\usepackage{stmaryrd}
\begin{document}

\preprint{??}

\title{Scaling in Counter Expressed Gene Networks Constructed from Gene
Expression Data}

\author{Himanshu \surname{Agrawal}}
\email{himanshu@mail.jnu.ac.in}
\affiliation{School of Information Technology,
	     Jawaharlal Nehru University, New Delhi -- 110067, India}

\date{\today}

\begin{abstract}

We study counter expressed gene networks constructed from gene-expression data
obtained from many types of cancers. The networks are synthesized by connecting
vertices belonging to each others' list of $K$-farthest-neighbors, with $K$
being an \textit{a priori} selected non-negative integer. In the range of $K$
corresponding to minimum homogeneity, the degree distribution of the networks
shows scaling. Clustering in these networks is smaller than that in equivalent
random graphs and remains zero till significantly large $K$. Their small
diameter, however, implies small-world behavior which is corroborated by their
eigenspectrum. We discuss implications of these findings in several contexts.

\end{abstract}


\maketitle

Gene networks are fundamental objects underlying the regulatory mechanism of
biological systems. Small portions of these networks have been studied by
molecular biologists since long using mutagenesis techniques pivoted on
manipulation of single gene, or at best a few genes, at a time \cite{Mutagen}.
Consequently, unraveling of gene networks has been slow. Recent advancements in
DNA microarray technology \cite{Microarray} have made large scale studies of
gene networks possible. Attempts have been made at identifying large segments of
regulatory networks using data from gene microarray expression assays
\cite{Network}. An aspect of these studies is identification of groups of
coregulated genes. This is done using specialized clustering techniques
developed in recent years \cite{Eis98Blat98Agr03}. Genes in each coexpression
group display similar expression pattern across different samples and are
expected to be coregulated. Identification of coexpressed genes is almost a
standard exercise in the expression profiling studies undertaken at present.

Regulatory genes function by means of both activating (up regulating) as well as
inhibiting (down regulating) the expression of genes. Knocking out of such a
gene leads to simultaneous change in the expression of genes that were either up
or down regulated by it. This may, in turn, result in a cascade affecting other
genes and destabilizing the organism. Thus, the coregulated genes are not
necessarily only coexpressed, they can also be counter expressed. Recent studies
have verified that both increase and decrease of expression level of genes are
equally discriminatory indicators of genetic pathologies \cite{Erns02}. Studies
on large scale properties of coexpressed gene networks constructed from gene
expression data have shown them to be having both small-world and scale-free
characteristic \cite{Agr02}. In this letter we analyze \textit{counter
expressed} \textit{gene networks} constructed from gene expression data and
outline results showing their relevance in biological processes.

Raw gene expression data requires extensive processing before it can be used
(see \cite{Preproc} for details). This gives an expression matrix having $N$
rows, each with zero mean and unit variance, corresponding to the $N$ genes and
$D$ columns corresponding to the samples. Henceforth this is the expression
matrix that we use and refer to. The algorithm for constructing counter
expressed gene networks is a modification of the algorithm used earlier for
constructing coexpressed gene networks \cite{Agr02}. It requires specification
of the maximum number of neighbors $K$, $0 \le K < N$, that a vertex can have.
For a given $K$, \textit{counter expressed} gene network is constructed using
the following two step procedure. (i) For each vertex $i$, $i = 1$, \ldots, $N$,
make a list $L_{i}$ of its $K$ farthest neighbors ordered by decreasing
distance. (ii) Connect all vertices $i$ and $j$ through an edge if $i \in L_{j}$
and $j \in L_{i}$, otherwise the vertices are not connected. We choose Euclidean
distance as the measure for making the list of farthest neighbors. The use of
other distance measures will not alter the results as long as they preserve the
ordering of points obtained from the Euclidean measure.

The topological structure of these networks, as in the case of coexpressed gene
networks \cite{Agr02}, is highly dependent on $K$. Starting with $N$ isolated
vertices at $K=0$, the network agglomerates very rapidly as $K$ is increased and
a single connected component is obtained for $K = K^{\star}$, $0 \ll K^{\star}/N
\ll 1$. A giant connected component emerges for $K = K_{\text{gcc}} \gtrsim 3$.
As this algorithm preferentially connects vertices that are far away from each
other, it selectively disfavors the formation of triangles. Consequently, we
expect that the clustering coefficient of these networks will be lower than that
of equivalent random graphs having the same number of vertices and edges. The
formation of squares, however, is preferred. The squares occur as long as there
are two groups of at least two vertices each such that the smallest distance
between vertices of different groups is larger than the largest distance between
vertices of the same group. The formation of squares, despite the absence of
triangles, makes the average length of shortest paths between vertices small and
gives these networks and their connected components a small-world structure.

\begin{figure}[t]
\centerline{\hskip-1mm%
	    \includegraphics[width=47mm]{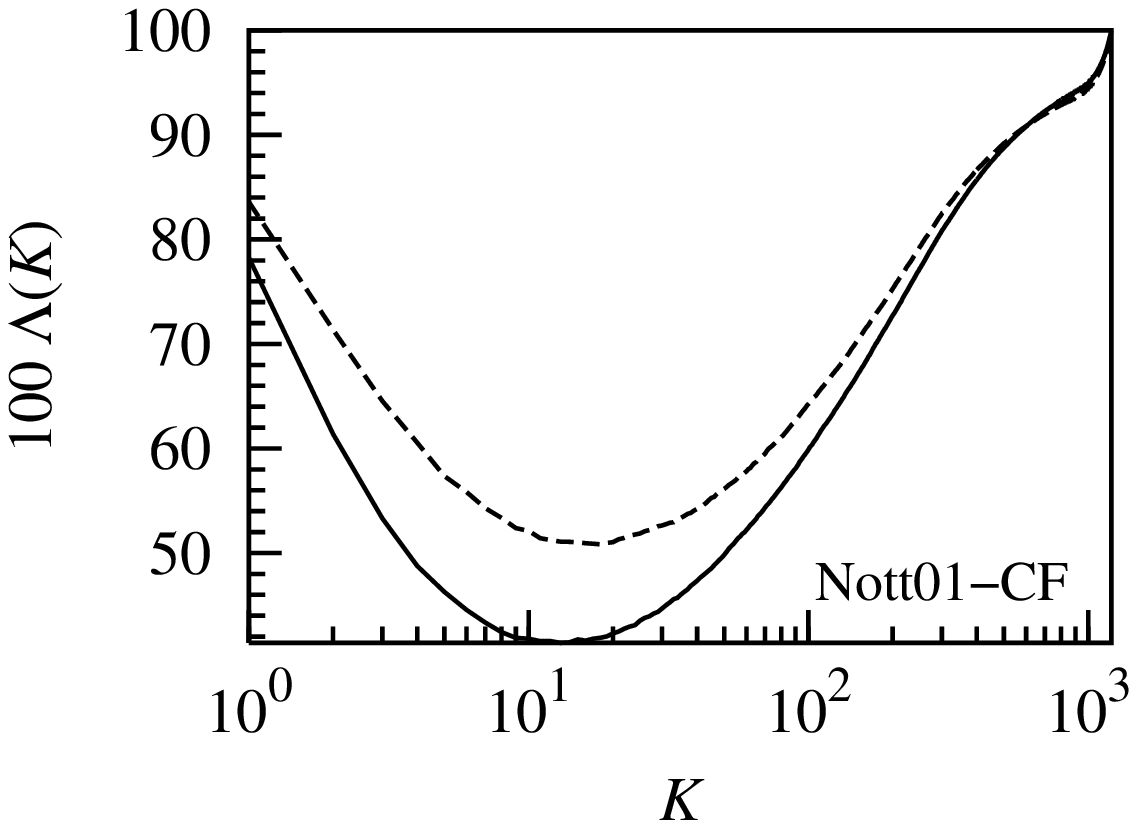}%
	    \hskip-2.5mm%
	    \includegraphics[width=47mm]{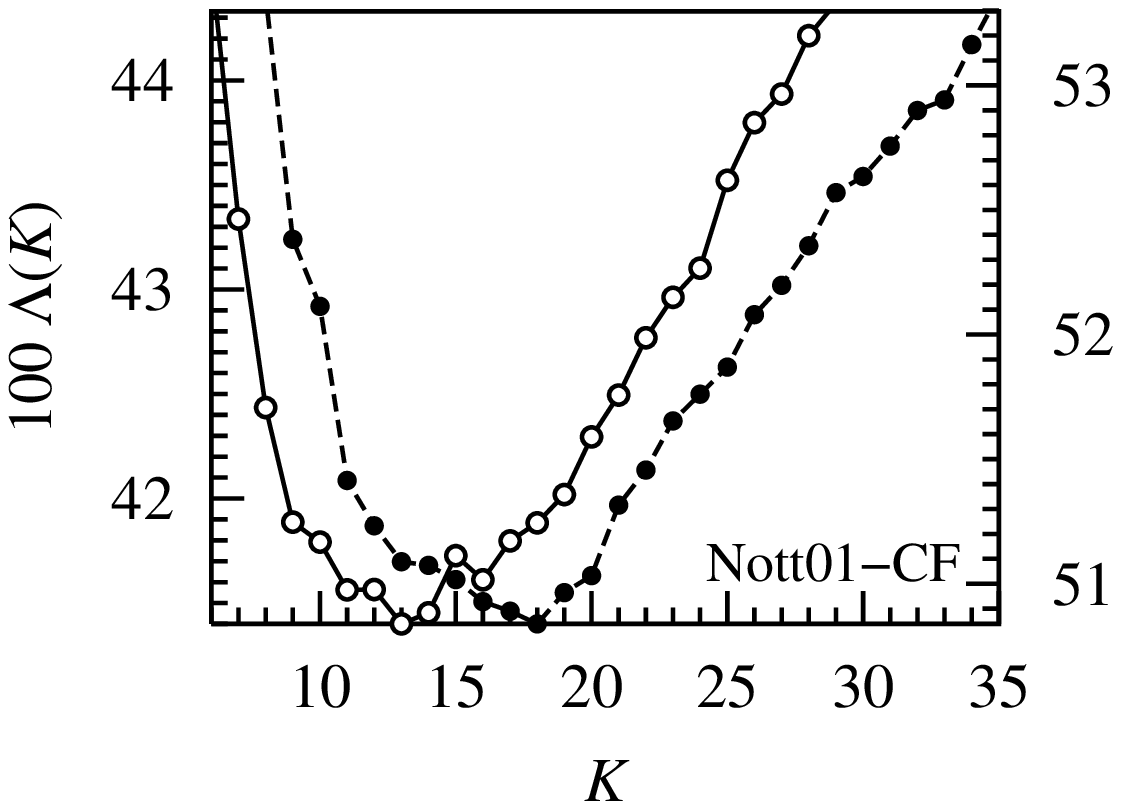}%
	    \hskip-2.5mm}

\caption{\label{F:LambK}Variation of the homogeneity $\Lambda(K)$ with $K$ in
counter expressed (solid lines) and coexpressed (dashed lines) gene networks
constructed from colon cancer data \cite{Nott01}. In the right subfigure the
solid line (with hollow circles) is on the left--bottom axes and the dashed line
(with solid circles) on the right--bottom axes. The circles mark the data
points.}
\end{figure}

\begin{figure}[b]
\centerline{\hskip-1mm%
	    \includegraphics[width=47mm]{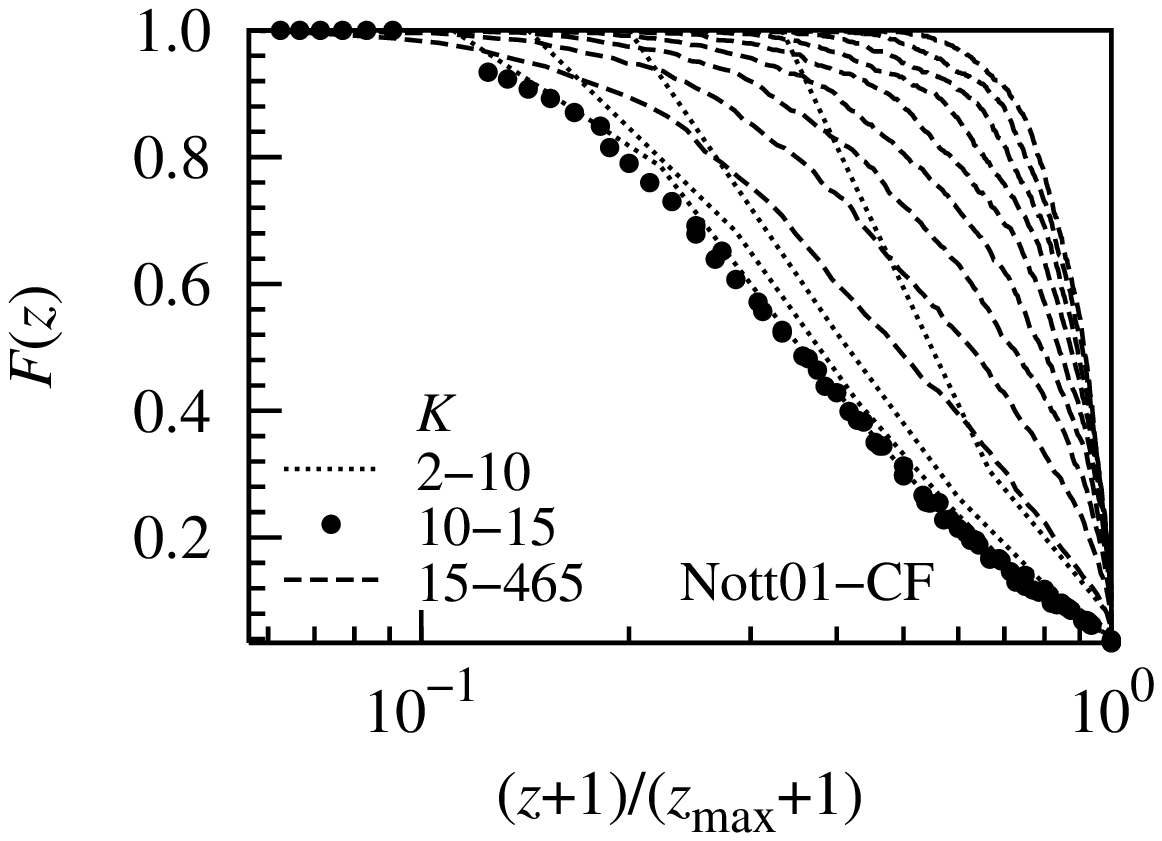}%
	    \hskip-2.5mm%
	    \includegraphics[width=47mm]{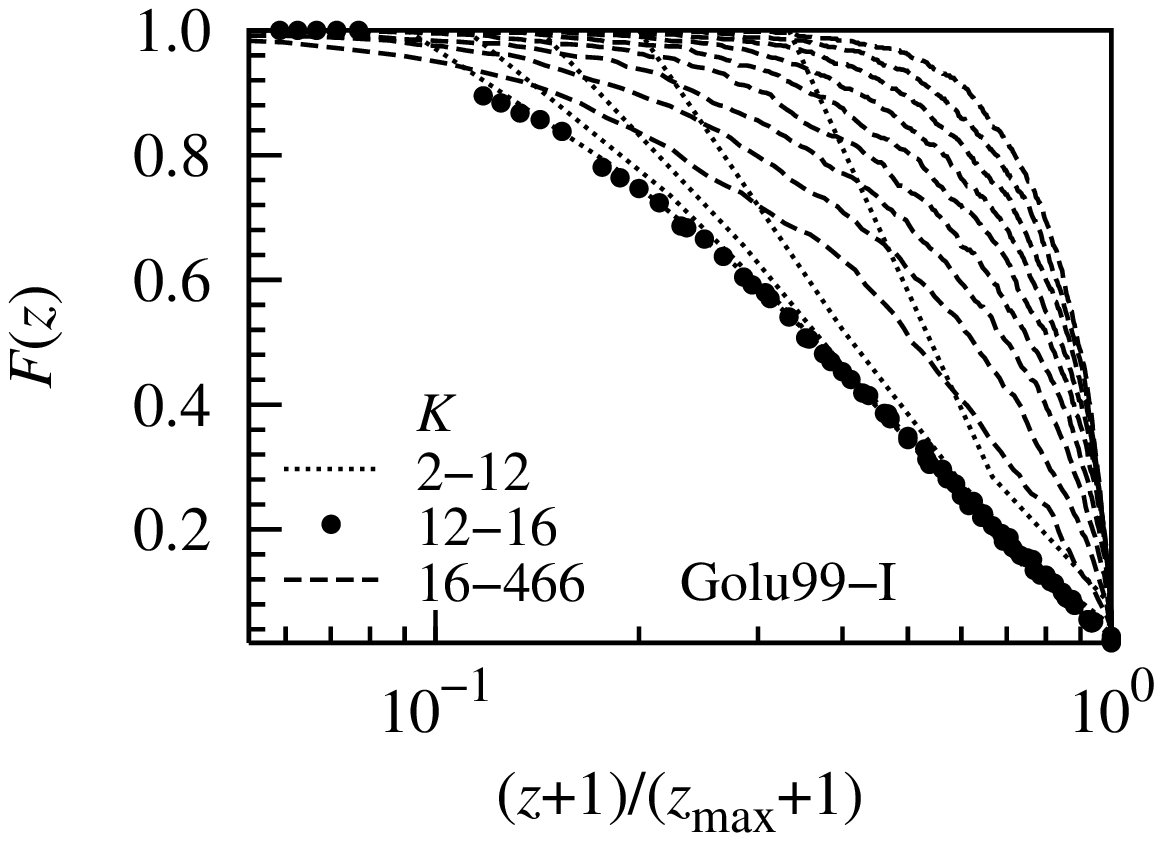}%
	    \hskip-2.5mm}
\centerline{\hskip-1mm%
	    \includegraphics[width=47mm]{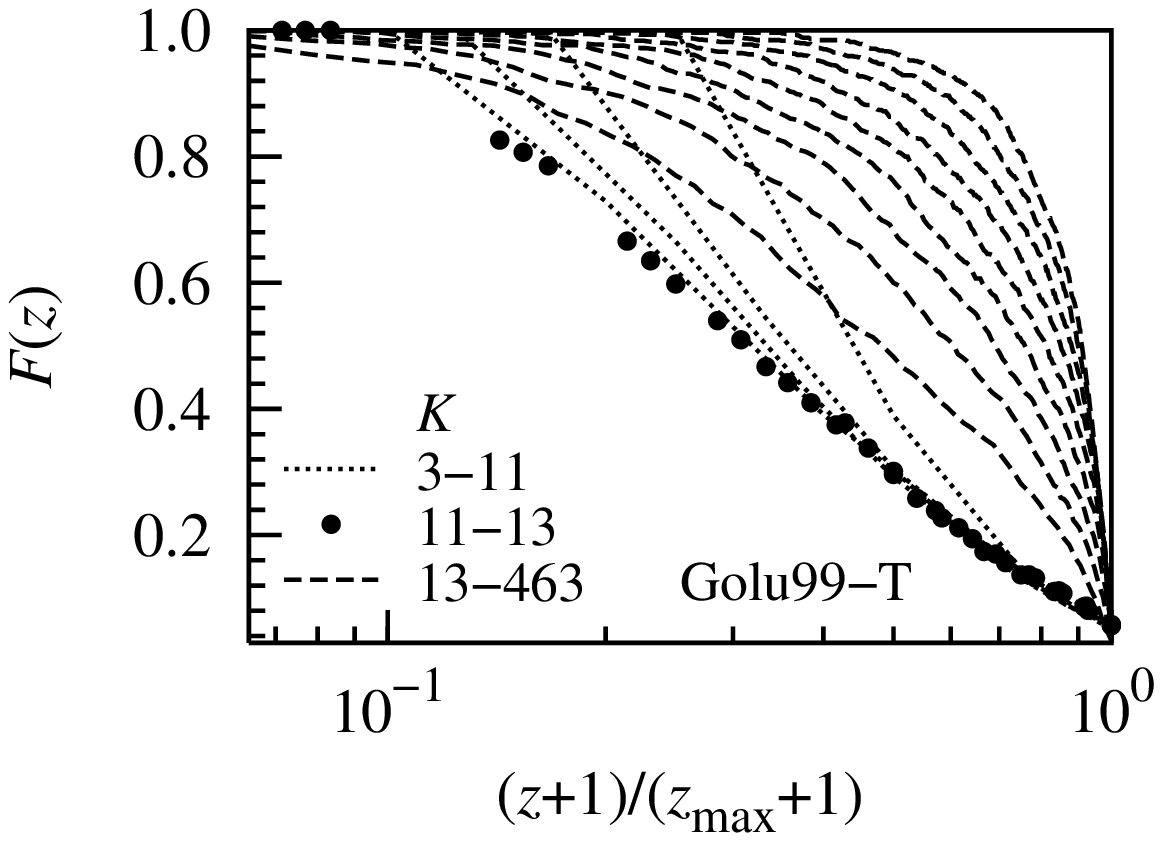}%
	    \hskip-2.5mm%
	    \includegraphics[width=47mm]{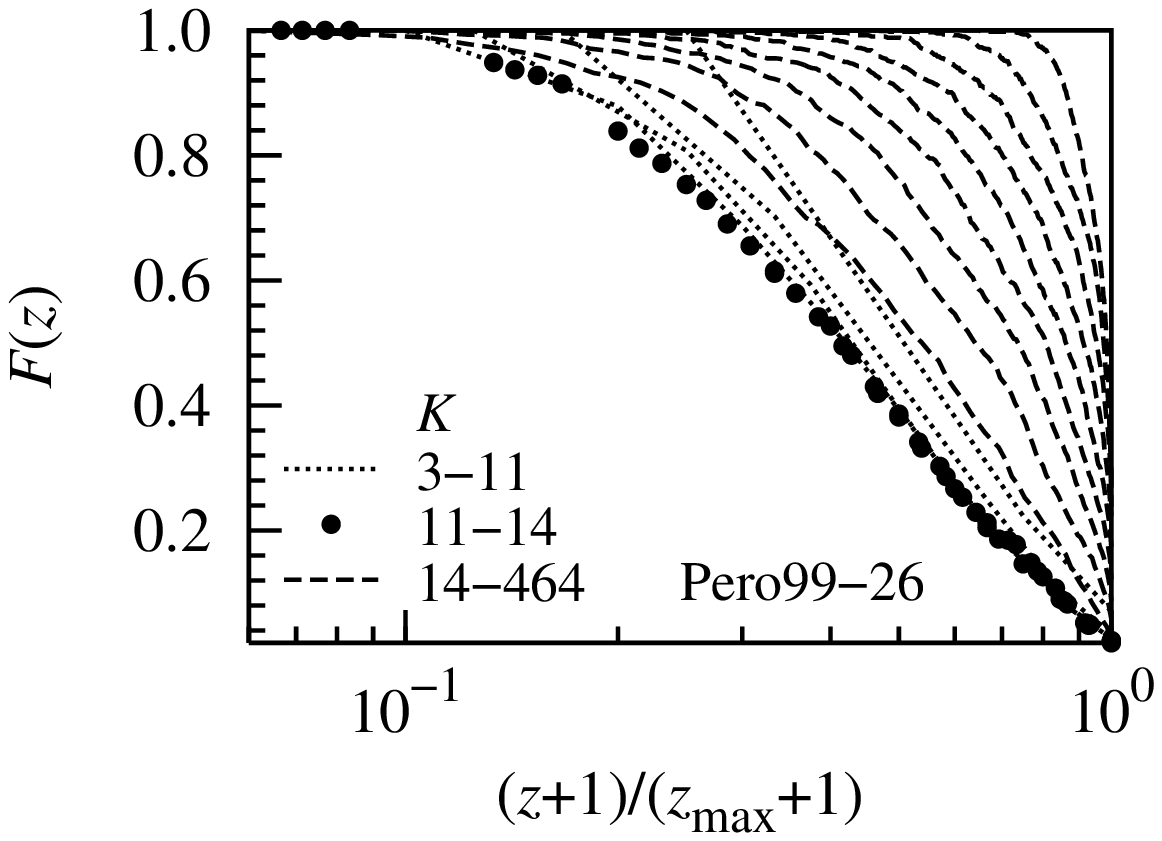}%
	    \hskip-2.5mm}

\caption{\label{F:DegDist}Variation of $F(z)$ with the normalized degree
$(z+1)/(z_{\text{max}}+1)$ in counter expressed networks constructed from
several gene expression data sets. Solid circles are used in the range of $K$
corresponding to minimum of $\Lambda(K)$. The dotted lines approach the solid
circles as $K$ increases (here, in steps of 2) and the dashed lines go away from
the solid circles as $K$ increases (here, in steps of 50). The keys are as in
\cite{Keys}.}
\end{figure}

We analyzed counter expressed networks constructed using gene expression data
from several types of cancers \cite{Nott01,Golu99,Pero99}. Let $z_{i}$ be the
degree of vertex $i$, $z_{\text{max}}$ be the largest degree in the network,
$P(z)$ be the degree density
\begin{equation}
  P(z) = \frac{1}{N} \sum_{i=1}^{N} \delta(z_{i}-z),
\end{equation}
and $F(z) = \sum_{i=z}^{z_{\text{max}}} P(i)$ be the distribution. For each
network we calculated $P(z)$, $F(z)$, and the homogeneity in terms of the
homogeneity order parameter $\Lambda(K)$ defined earlier \cite{Agr02}. The
variation of homogeneity in both coexpressed and counter expressed gene networks
constructed from colon cancer data \cite{Nott01} is shown in Fig.~\ref{F:LambK}.
The figure clearly shows that on increasing $K$ the homogeneity of both the
types of networks first decreases for small $K$, reaches a minimum, and then
increases to its maximum value of $1$ at $K = N-1$. The minimum, as seen in the
Fig.~\ref{F:LambK}, is somewhat noisy and flat in both the networks.
Furthermore, for the same $K$ the homogeneity of the counter expressed networks
is smaller than that of the coexpressed networks, except at $K$ of the order of
$N$.

Figure~\ref{F:DegDist} shows the variation of the observed cumulative
probability distribution function $F(z)$ with normalized degree
$(z+1)/(z_{\text{max}}+1)$ in a wide range of values of $K$ for counter
expressed networks constructed from several gene expression data sets. From the
figure it is clear that in the range $K_{1} \le K \le K_{2}$ of $K$
corresponding to the flat minimum of the order parameter $\Lambda(K)$, the
degree distributions of the networks collapse together and show good scaling for
all the data sets. The tails of these distributions fit well with the form
\begin{eqnarray}\label{E:Fz-z}
  F(z) = a-b\ln\left(\frac{z+1}{z_{\text{max}}+1}\right),
\end{eqnarray}
where $a$ and $b$ are fit parameters. This implies that the corresponding
$P(z)$ has scale-free behavior of the form $b(z+1)^{-1}$ in
the tails, sharply truncated at $z = z_{\text{max}}$.
As the number of hubs $\mathfrak{H}$ in these networks varies in the range
$20$--$60$, the observed $z_{\text{max}}$ is consistent with that expected
$N/\mathfrak{H} H(z_{\text{max}}+1)-1$ for sharply truncated power-law
probability density with exponent $-1$, where $H(n)$ is $n$th Harmonic number.

\begin{figure}
\centerline{\hskip-1mm%
	    \includegraphics[width=47mm]{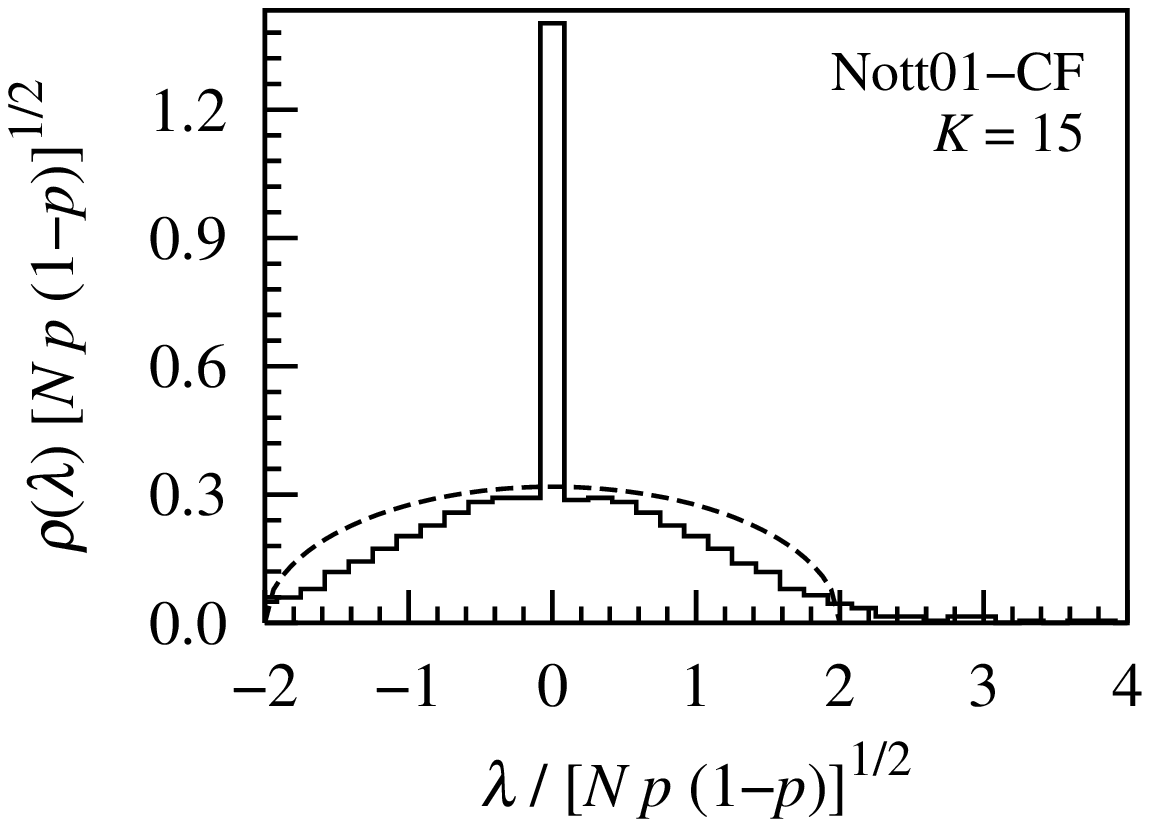}%
	    \hskip-2.5mm%
	    \includegraphics[width=47mm]{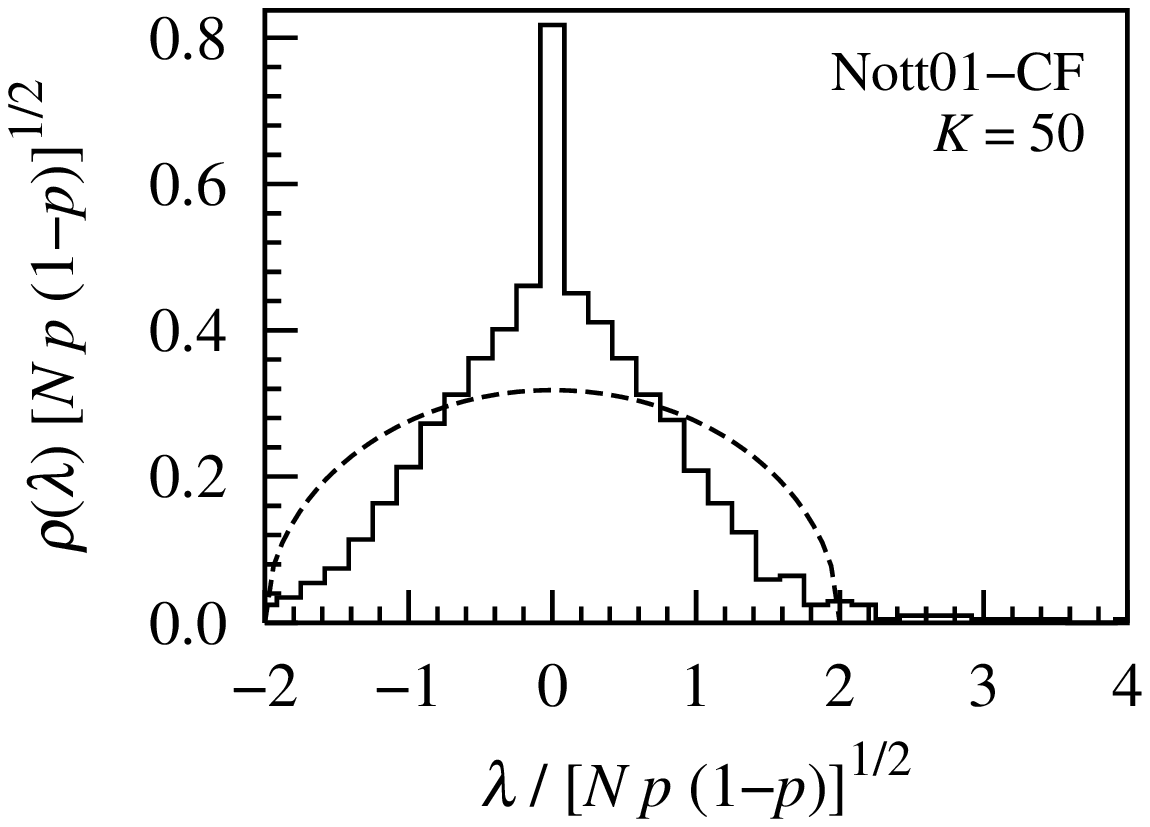}%
	    \hskip-2.5mm}
\centerline{\hskip-1mm%
	    \includegraphics[width=47mm]{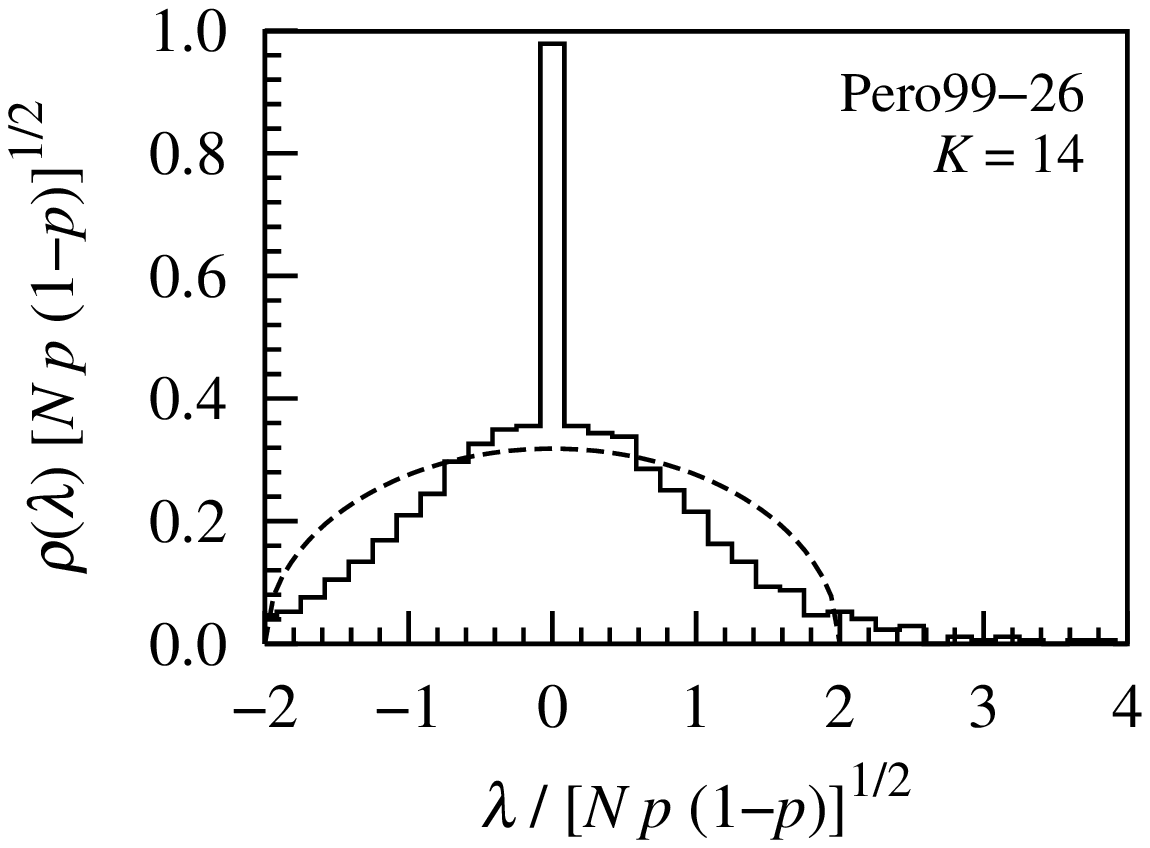}%
	    \hskip-2.5mm%
	    \includegraphics[width=47mm]{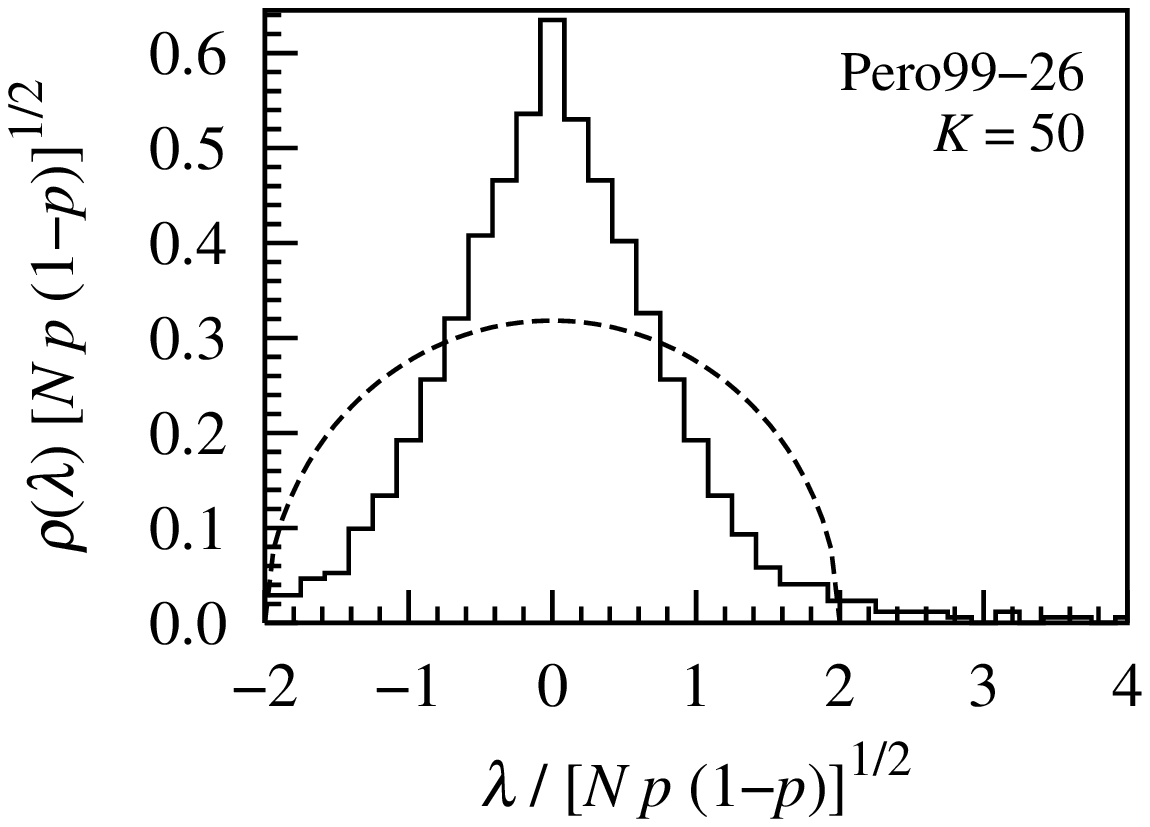}%
	    \hskip-2.5mm}

\caption{\label{F:SpecDens}Spectral density $\rho(\lambda)$ of counter expressed
gene networks constructed from several gene expression data sets. Semicircle
corresponding to spectral density of random networks is drawn for comparison.
The keys are as in \cite{Keys}.}
\end{figure}

The scale-free behavior, seen above in $50$--$75\%$ of the range of variation of
the normalized degree, is also indicated by the spectral density $\rho(\lambda)$
of the eigenvalue spectrum of the adjacency matrix of the networks
\cite{Far01Goh01}
\begin{equation}\label{E:RhoLamb}
  \rho(\lambda) = \frac{1}{N} \sum_{j=1}^{N} \delta(\lambda-\lambda_{j}),
\end{equation}
where $\lambda_{j}$ is the $j$th eigenvalue. Figure~\ref{F:SpecDens} shows that
$\rho(\lambda)$ develops a triangular form for $K \ge K_{1}$, indicating the
presence of a power-law in the networks. The triangular shape persists till $K
\approx N-1$. For $K < K_{1}$, $\rho(\lambda)$ is highly skewed with several
blurred peaks indicating the presence of small-world behavior. Similar behavior
observed earlier in coexpressed gene networks \cite{Agr02} was accompanied by
very high clustering coefficient. In the present case, however, the clustering
coefficient has a completely different behavior. This implies that the
structural details of counter and coexpressed networks, contrary to the
intuition, are very different.

\begin{figure}[t]
\centerline{\hskip-1mm%
	    \includegraphics[width=47mm]{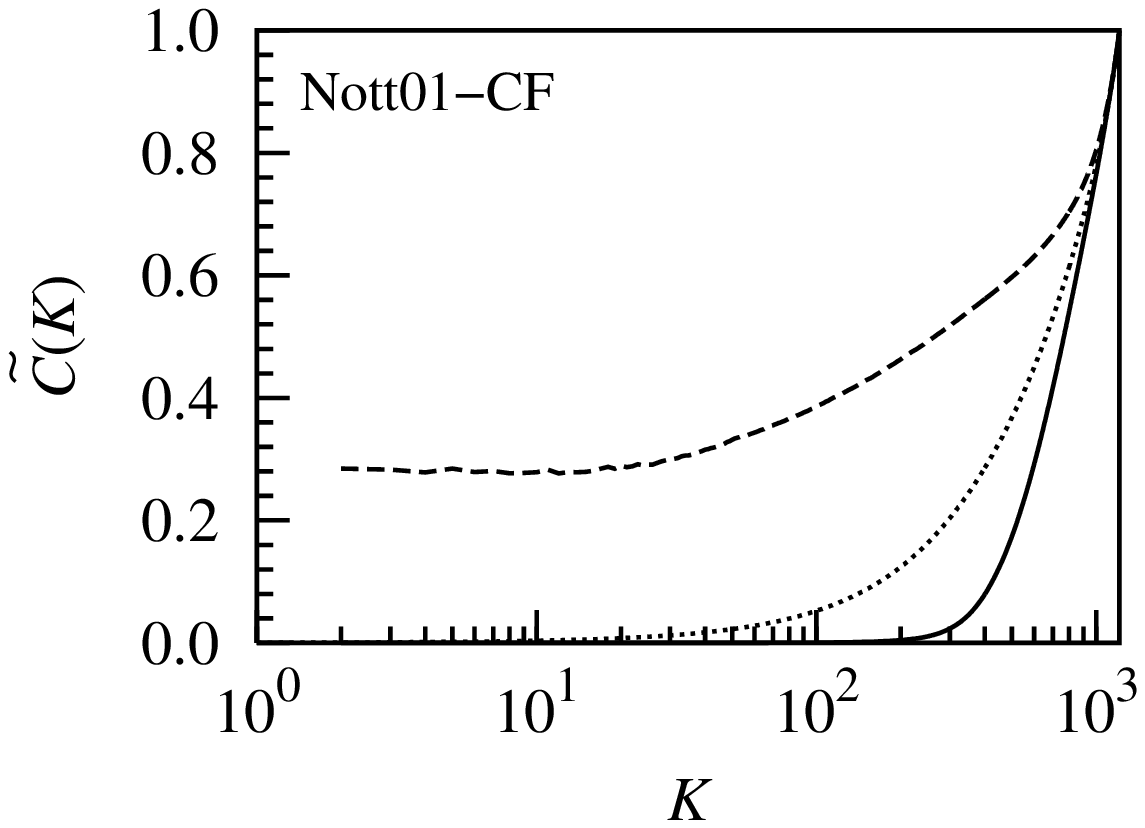}%
	    \hskip-2.5mm%
	    \includegraphics[width=47mm]{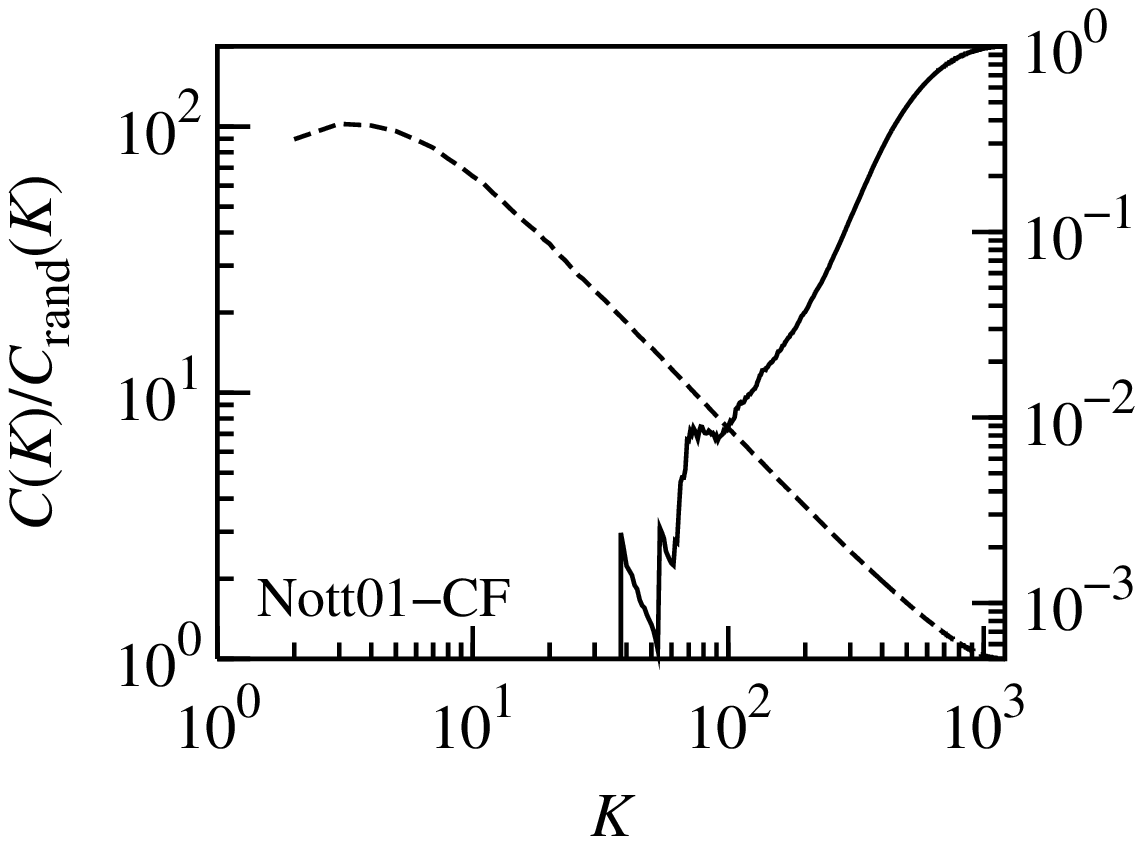}%
	    \hskip-2.5mm}

\caption{\label{F:ClustK}Variation of $\widetilde{C}(K)$ and
$C(K)/C_{\text{rand}}(K)$ with $K$ in counter expressed (solid lines) and
coexpressed (dashed lines) gene networks constructed from colon cancer data
\cite{Nott01}. The dotted line corresponds to random graphs equivalent to the
coexpressed (also indistinguishably equivalent to the counter expressed) gene
networks. In the right subfigure the solid (dashed) line is on the right
(left)--bottom axes.}
\end{figure}

The clustering coefficient $C_{i}(K)$ of a vertex $i$ of degree $z_{i}$ is
defined as
\begin{equation}\label{E:CCvert}
  C_{i}(K) =
    \left\{
      \begin{array}{ll}
	0                              &, z_{i} \le 1 \\
	\displaystyle
	\frac{y_{i}}{z_{i}(z_{i}-1)/2} &, z_{i} \ge 2
      \end{array}
    \right.,
\end{equation}
where $y_{i}$ is the total number of links present among the $z_{i}$ neighbors
of the vertex. The clustering coefficient of the network is defined as the mean
of $C_{i}(K)$ in two different ways. First, it is $C(K)$ the average over
\textit{all} the vertices. Second, it is $\widetilde{C}(K)$ the average taken
only over vertices of degree at least 2. The clustering coefficient of the
equivalent random graph is $C_{\text{rand}}(K) = E/[N(N-1)/2]$, where $E$ is the
number of edges in the graph \cite{Watt98}.

The variation of the clustering coefficient with $K$ for both coexpressed and
counter expressed gene networks is shown in Fig.~\ref{F:ClustK}. The figure
clearly shows that $\widetilde{C}(K)$ of coexpressed networks is much higher
than that of equivalent random graphs for all $K$. It is almost constant at
$\approx 0.277$ for small $K$ and then increases to its maximum values of 1 as
$K$ increases to $N-1$. The $C(K)$ also has similar behavior except at small $K$
when it increases rapidly from $\approx 0.055$ at $K = 2$ to $\approx 0.229$ at
$K = 10$ and then to $\approx 0.277$ at $K = 20$. Despite this $C(K)$ remains
higher than that for the corresponding random graphs for all $K$, except at $K =
N-1$ when both are 1, as seen by the variation of $C(K)/C_{\text{rand}}(K)$ in
Fig.~\ref{F:ClustK}.

Figure~\ref{F:ClustK} shows that the clustering coefficient of counter expressed
gene networks remains smaller than that for the corresponding random graphs for
all $K$. Both become equal and equal to their maximum value of 1 only at $K =
N-1$. Furthermore, both $C(K)$ and $\widetilde{C}(K)$ remain zero till
significantly large $K = K_{\triangle}$ in these networks (here $K_{\triangle} =
37$). This is a consequence of absence of triangles. Selective disfavoring of
triangle formation, in general, leads to $C(K)/C_{\text{rand}}(K) < 1$ for all
$K < N-1$.

\begin{figure}
\centerline{\hskip-1mm%
	    \includegraphics[width=47mm]{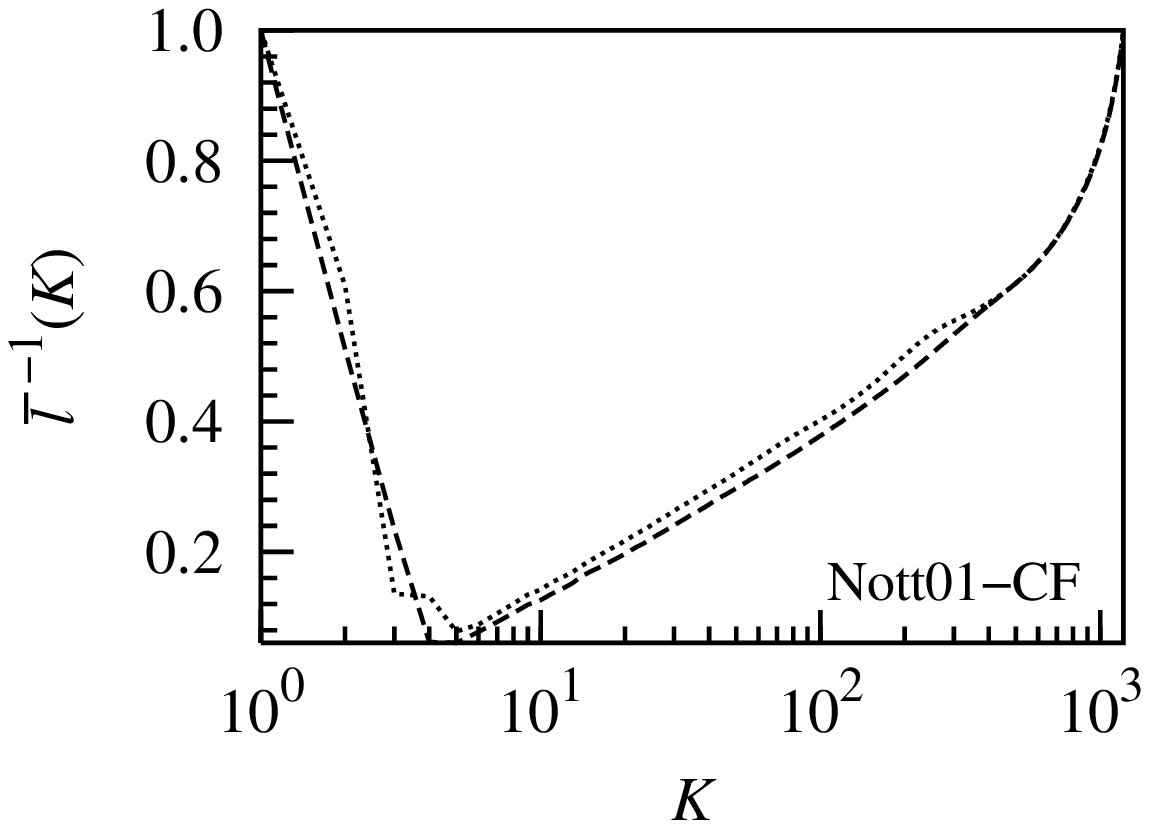}%
	    \hskip-2.5mm%
            \includegraphics[width=47mm]{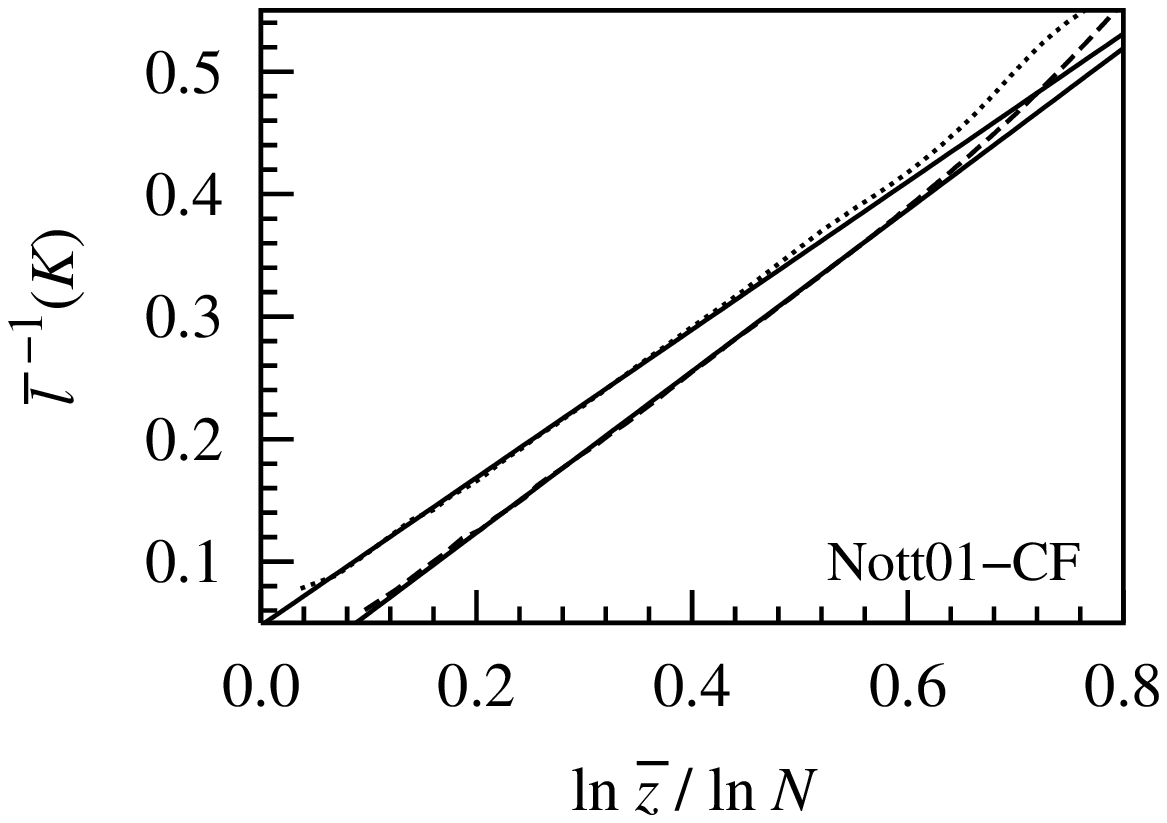}%
	    \hskip-2.5mm}

\caption{\label{F:MeanSP}Left panel: Variation of $\bar{\ell}(K)$ the average
shortest path length between vertices with K in counter expressed (dotted lines)
and coexpressed (dashed lines) gene networks constructed from colon cancer data
\cite{Nott01}. Right panel: Interdependence between $\bar{\ell}(K)$ and the mean
degree $\bar{z}(K)$ of vertices. The solid lines are fitting straight lines.}
\end{figure}

The most important indicator of small-world behavior is the mean $\bar{\ell}$
and the maximum $\ell_{\text{max}}$ of the shortest paths between mutually
reachable vertices. In disjoint networks these are defined by using only the
finite length shortest paths. In all networks having $N$ vertices, $\bar{\ell}$
and $\ell_{\text{max}}$ are bounded in $[1, (N+1)/3]$ and $[1, N-1]$,
respectively. The variation of $\bar{\ell}$ with $K$ in both counter expressed
and coexpressed gene networks is shown in Fig.~\ref{F:MeanSP}. The figure shows
that in both types of networks $\bar{\ell}^{-1}$, starting with 1 at $K = 1$,
decreases sharply as $K$ increases till $K = K_{\text{gcc}}$, the value at which
giant connected component emerges in the network. On increasing $K$ further,
$\bar{\ell}^{-1}$ decreases and attains a minimum at $K = K_{\merge} \ge
K_{\text{gcc}}$. This occurs because large chunks merge, forming still larger
chunks and introducing paths, longer than $\bar{\ell}$, connecting vertices
across the merged chunks. At $K = K_{\merge}$ the effect of long paths
introduced by merging of chunks is balanced by the simultaneously introduced
short paths and reduction in the length (if any) of existing paths due to new
shortcuts within the chunks. Increasing $K$ beyond $K_{\merge}$ leads to
decrease in $\bar{\ell}$, as seen by the behavior of $\bar{\ell}^{-1}$ in
Fig.~\ref{F:MeanSP}, till the vertices form a complete graph. In data containing
widely separated big chunks, merging and consequent rapid increase in
$\bar{\ell}$ can occur at $K > K_{\merge}$. This will appear as sharp dips in
the presently smooth $\bar{\ell}^{-1}$ versus $K$ curve seen in
Fig.~\ref{F:MeanSP}. The figure shows that immediately after the minimum,
$\bar{\ell}^{-1}$ follows $\hat{a}+\hat{b}\ln\bar{z}/\ln N$ very well till the
onset of finite size effects at large $K$, where $\bar{z}$ is mean degree. This
behavior is characteristic of both random graphs and small-world networks
\cite{Alb02Dor02}. The behavior of $\ell_{\text{max}}$, although more noisy, was
similar to that of $\bar{\ell}$.

\begin{figure}[t]
\centerline{\hskip-1mm%
	    \includegraphics[width=47mm]{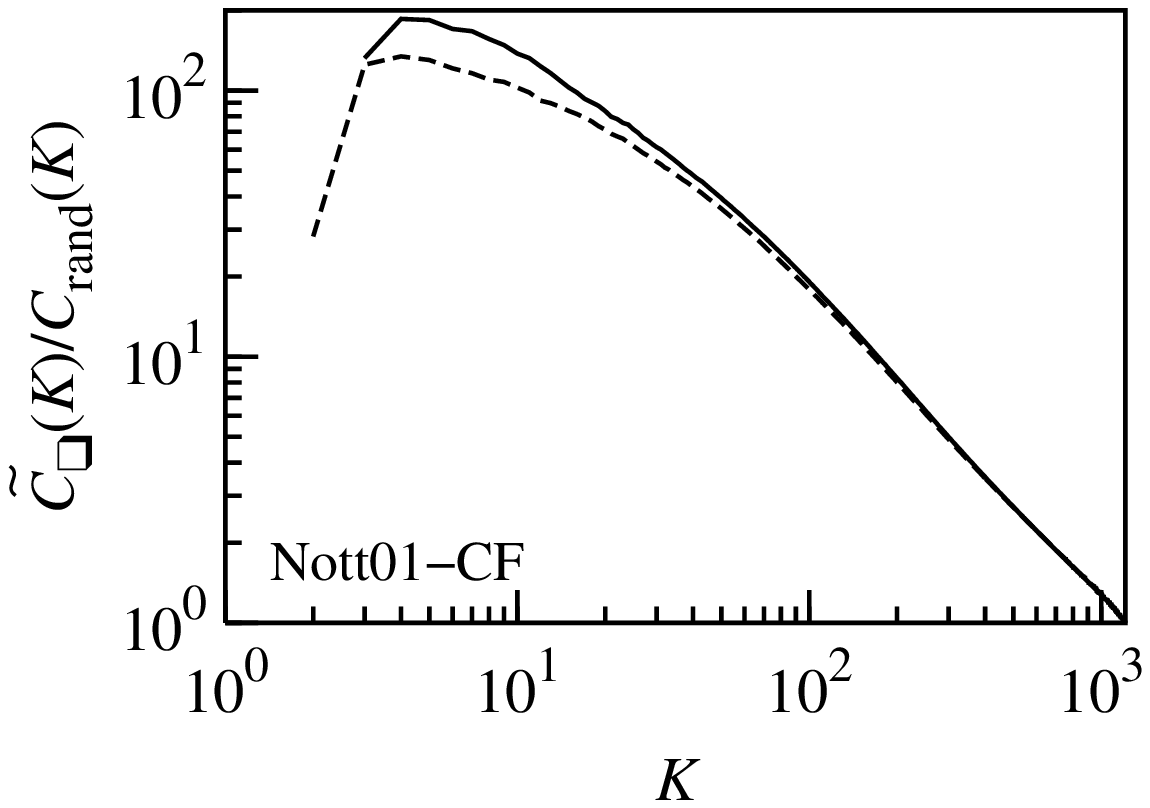}%
	    \hskip-2.5mm%
	    \includegraphics[width=47mm]{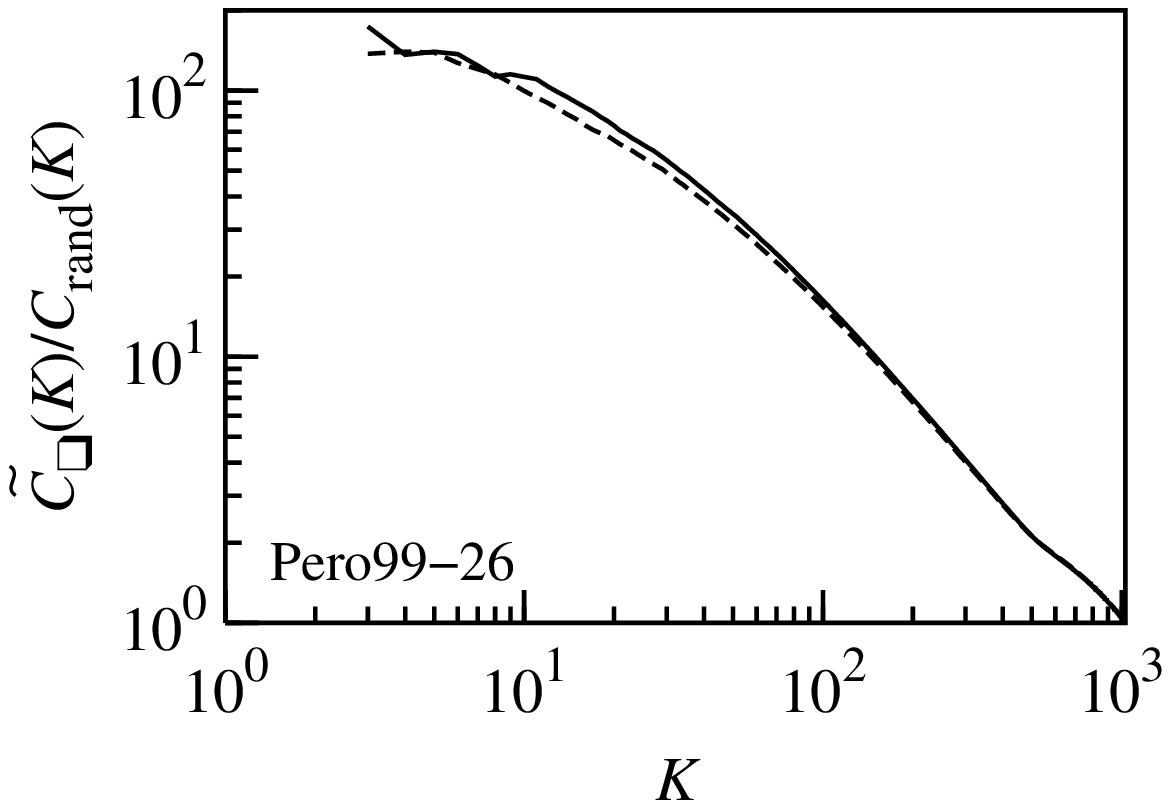}%
	    \hskip-2.5mm}

\caption{\label{F:Clust4K}Variation of
$\widetilde{C}_{\square}(K)/C_{\text{rand}}(K)$ with $K$ in counter expressed
(solid lines) and coexpressed (dashed lines) gene networks from several data
sets. The keys are as in \cite{Keys}.}
\end{figure}

Zero clustering for $K \le K_{\triangle}$ and very low clustering otherwise
observed in counter expressed networks is a consequence of the definition of
clustering coefficient which looks for triangles in the network. This
definition, although good for random graphs, does not always reveal the actual
extent of clustering. We defined new clustering coefficient using squares
instead of triangles by setting $y_{i}$ in Eq.~(\ref{E:CCvert}) to the number of
pairs of neighbors of $i$ that have at least one more common neighbor. The
square based clustering coefficients $\widetilde{C}_{\square}$ and
$C_{\square}$, computed similarly to $\widetilde{C}$ and $C$, respectively, have
much higher values relative to random graphs for both counter expressed and
coexpressed networks. The $C_{\square,\text{rand}}$ value for random graphs is
the probability that two vertices connected to a vertex (say, ``$\text{O}$'')
have another common neighbor (say, ``$\Theta$''; $\Theta \not\equiv \text{O}$).
It evaluates to $1-(1-p)^{\bar{z}-1}$ for large $N$, where $p=\bar{z}/(N-1)$ is
the edge density. The variation of
$\widetilde{C}_{\square}(K)/C_{\text{rand}}(K)$ is shown in
Fig.~\ref{F:Clust4K}. The behavior of $C_{\square}(K)/C_{\text{rand}}(K)$ is
similar. These results show that clustering in counter expressed networks,
despite $C(K)$ being zero, is actually very high and comparable to that in
coexpressed networks.

An implication of $C(K) = 0$ for $K \le K_{\triangle}$ is that all the connected
components comprising the counter expressed networks are bipartite for these
values of $K$. This allows each connected component to be partitioned into two
groups such that the genes within each group are coexpressed but those in
different groups are counter expressed. This is done by assigning a gene to one
group (say, A) and its neighbors to the other group (say, B) and then
iteratively assigning neighbors of genes in group A (or, B) to group B (or, A)
till no gene in the connected component remains unassigned. This procedure can
be used even if the network contains triangles by first eliminating them, e.g.,
in the counter expressed networks for $K > K_{\triangle}$, by iteratively
identifying all the links forming triangles and removing the smallest one.
Triangle elimination may decompose a connected component into two or more parts.
In such a case, isolated vertices, doublets, and small chunks are ignored and
each of the big chunks is separately processed further. However, the larger the
number of triangles that need to be removed before partitioning, the smaller
will be the clarity of partitions.

The results presented above were verified using counter expressed networks
constructed from several other gene expression data sets also \cite{OtherData}.
Flat minimum of $\Lambda(K)$ implies that counter expressed gene networks have
non-trivial structure that is robust against noise. High $C_{\square}(K)$ and
$\widetilde{C}_{\square}(K)$ imply that the structure is highly clustered. Thus,
these networks can be used for non-parametric partitioning of gene expression
data. The scale-free character of these networks implies that they are built
around few crowded hubs and are robust. The small-world structure, a key
evolutionary aspect, implies that signaling mechanism encoded by these networks,
despite the lack of triangles, functions as efficiently as that of coexpressed
networks. These results show that the role of counter expressed networks is as
important in biological processes as that of coexpressed networks. Thus, despite
difference in their regulatory behavior (activating versus inhibiting gene
expression), both these types of networks should be considered together.

\end{document}